\documentclass[aps,prb,twocolumn,superscriptaddress,showpacs,preprintnumbers]{revtex4}

\usepackage{epsfig}
\usepackage{tikz}
\usepackage{graphicx}

\usepackage{color}
\usepackage[colorlinks,bookmarks=false,citecolor=blue,linkcolor=red,urlcolor=blue]{hyperref}
\usepackage{dcolumn}
\usepackage{bm}

\usepackage{amsmath,amssymb}

\begin{document}

\title{Impurity induced current in a Chern insulator}

\author{Vibhuti Bhushan Jha${\footnote[1]{These authors contributed equally to this work.}}$ }
\affiliation{The Institute of Mathematical Sciences, C I T Campus, Chennai 600 113, India}
\affiliation{Indian Institute of Space Science and Technology,  Valiamala Road, Valiamala, Thiruvananthapuram, Kerala 695547, India}
\author{Garima Rani$^*$}
\affiliation{The Institute of Mathematical Sciences, C I T Campus, Chennai 600 113, India}
\affiliation{Homi Bhabha National Institute, Training School Complex, Anushakti Nagar, Mumbai 400094, India}
\author{R. Ganesh}
\affiliation{The Institute of Mathematical Sciences, C I T Campus, Chennai 600 113, India}
\affiliation{Homi Bhabha National Institute, Training School Complex, Anushakti Nagar, Mumbai 400094, India}

\date{\today}

\begin{abstract}
Chern insulators arguably provide the simplest examples of topological phases. They are characterized by a topological invariant and can be identified by the presence of protected edge states. In this article, we show that a local impurity in a Chern insulator induces a twofold response: bound states that carry a chiral current  
and a net current circulating around the impurity. This is a manifestation of broken time 
 symmetry and persists even for an infinitesimal impurity potential. To illustrate this, we consider a Coulomb impurity in the Haldane model. Working in the low-energy long-wavelength limit, we show that an infinitesimal impurity strength suffices to create bound states. We find analytic wavefunctions for the bound states and show that they carry a circulating current. In contrast, in the case of a trivial analogue -- graphene with a gap induced by a sublattice potential -- bound states  occur but carry no current. 
In the many body problem of the Haldane model at half-filling, we use a linear response approach to demonstrate a circulating current around the impurity. 
Impurity textures in insulators are generally expected to decay exponentially; in contrast, this current decays polynomially with distance from the impurity. 
Going beyond the Haldane model, we consider the case of coexisting trivial and non-trivial masses. We find that the impurity induces a local chiral current as long as time reversal symmetry is broken. However, the decay of this local current bears a signature of the overall topology -- the current decays polynomially in a non-trivial system and exponentially in a trivial system. 
In all cases, our analytic results agree well with numerical tight-binding simulations.
\end{abstract}
\keywords{}
\maketitle

\section{Introduction}
Chern insulators are lattice analogues of the integer quantum Hall effect in which hopping terms break time reversal symmetry. The first Chern insulator was proposed by Haldane in 1988\cite{Haldane1988}, providing the starting point for the physics of topological insulators.
The classic signature of a Chern insulator is the occurrence of chiral edge states\cite{Halperin1982,Hao2008,Mong2011}. 
Our main result -- impurity-induced chiral current -- can be thought of as a limiting case of an edge current around a hole cut into the sample. Its closed edge will carry edge states and a net chiral current. As the hole is shrunk to a point, we are left with a localized impurity potential. The chiral edge current maps to a current that circles the impurity. Furthermore, edge states at the edge of the hole reduce to impurity-centred bound states which carry persistent chiral currents.
We demonstrate this in the Haldane model with a Coulomb impurity potential, chosen for ease of analytic calculations. 

The early work of Prange\cite{Prange1981} pointed out that impurities in quantum Hall systems lead to localized bound states, but do not affect the macroscopic Hall current.
More recently, general considerations have been laid out for the occurrence of impurity bound states in topological systems\cite{Slager2015,Kimme2016}. Here, we show that an impurity in a Chern insulator leads to a non-trivial texture carrying persistent chiral currents. Similar results have been reported with single-site impurities in the Kane Mele model\cite{Gonzalez2012}.

We build upon several studies of Coulomb impurities in Dirac systems which have largely been motivated by graphene\cite{Pereira2007,Biswas2007,Novikov2007,Terekhov2008,Gupta2010}.
 Here, we focus on gapped Dirac Hamiltonians\cite{Pereira2008,Gupta2008,Zhu2009,Skinner2014}. We consider two ways of opening an insulating gap on the honeycomb lattice: (a) by a sublattice potential\cite{Semenoff1984,Hunt2013}, henceforth referred to as the `trivial' case, and (b) by a complex next-nearest neighbour hopping -- henceforth, the `non-trivial' case. 
While the former breaks inversion symmetry, the latter breaks time reversal symmetry and leads to a Chern insulator\cite{Haldane1988}. We first consider each case separately and show that chiral currents only emerge in the non-trivial case. We later consider the general case in which both these effects coexist. 

  \begin{figure}
  \includegraphics[width=2.in]{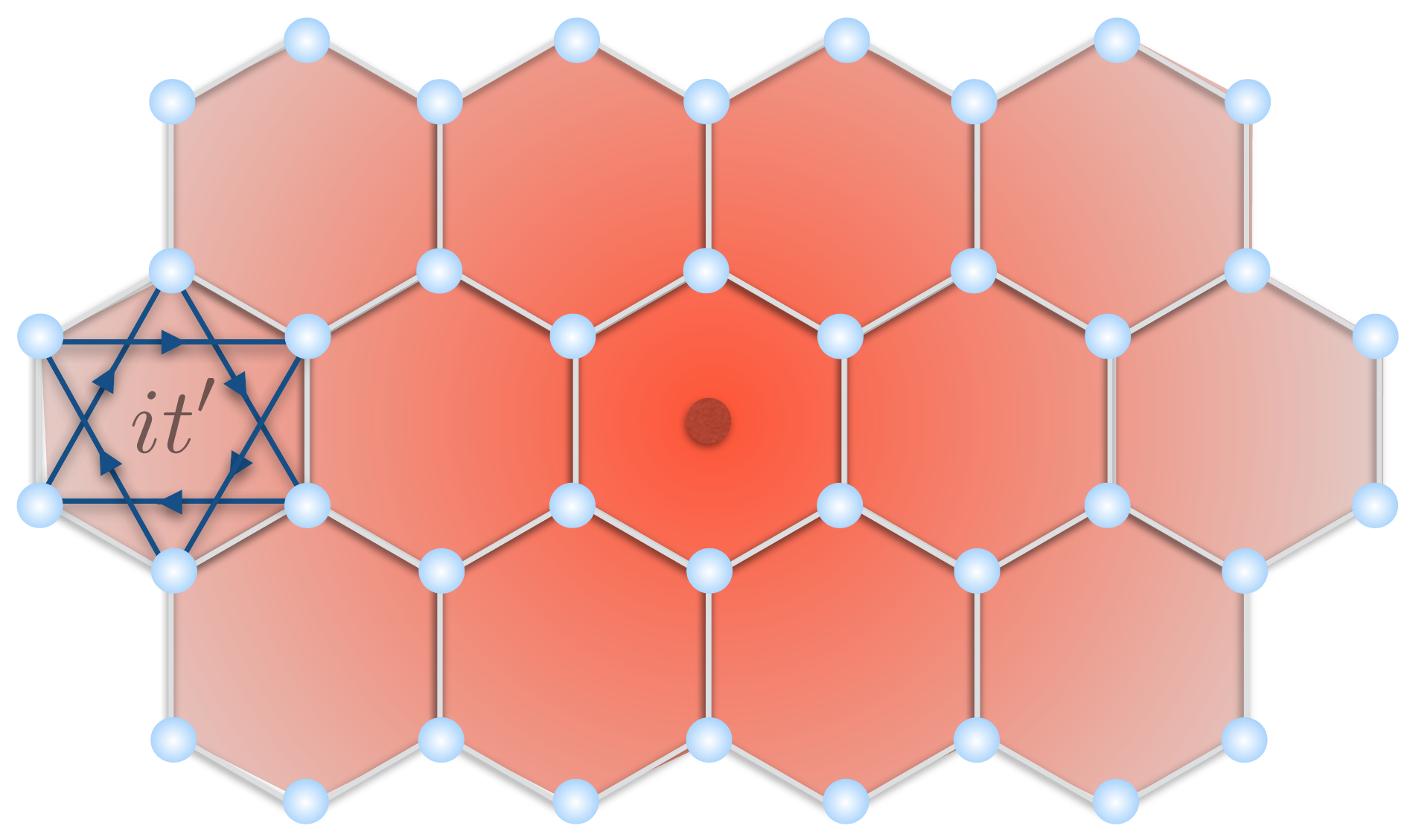}
  \caption{Honeycomb lattice with a Coulomb impurity. The complex nearest-neighbour hopping of the Haldane model is shown in one plaquette. }
  \label{fig.honeycomb}
 \end{figure}

\section{Coulomb impurity in a Dirac system}
We study spinless fermions on a honeycomb lattice described by the Hamiltonian 
\begin{eqnarray} 
H = -t \sum_{\langle ij \rangle} \{ c_{i}^\dagger c_j + h.c.\}  + H_{mass} + \sum_i \frac{g}{\vert \mathbf{r_i} - \mathbf{r}_0 \vert} c_{i}^\dagger c_{i}.
\end{eqnarray} 
The first term represents nearest neighbour hopping as in graphene. The last term represents a repulsive Coulomb impurity located at the centre of one hexagon as shown in Fig 1. The term in the middle represents a mass term that opens a band gap, which may be of two types
\begin{eqnarray}
H_{mass,trival} &=& \sum_{i\in A} V c_{i}^\dagger c_{i} - \sum_{i\in B} V c_{i}^\dagger c_i, \\
H_{mass,non-trivial} &=& it'  \sum_{\langle \langle ij \rangle \rangle} \{ c_{i}^\dagger c_{j} - h.c. \}.
\end{eqnarray} 
The numerical results presented below are obtained from the diagonalization of this Hamiltonian with the appropriate mass term implemented as above with  $V/t = 0.1$ for the trivial case and $t'/t = 0.1$ for the non-trivial case. The tight binding calculations have been performed on clusters of 7200 sites. We have assumed periodic boundary conditions to avoid edge states, thereby allowing us to easily identify states within the band gap as bound states. The Coulomb impurity potential decays rapidly with distance; we cut it off at the edge of the cluster so as to not leak into neighbouring clusters. Except in extreme cases ( $g \ll t$), the bound states that we find are also well localized within the cluster.

In the absence of impurities, this Hamiltonian leads to two gapped Dirac cones at the $\mathbf{K}$-points of the Brillouin zone. 
The long-wavelength response to an impurity is then determined by 
states near the two $\mathbf{K}$ points, described by the Hamiltonians\cite{Pereira2007}
\begin{eqnarray}
\nonumber \mathcal{H}_{\mathbf{K}}= \left( \begin{array}{cc}
 -\lambda_{\mathbf{K}}+g/r & \alpha e^{-i\theta}(\partial_r-\frac{i}{r}\partial_{\theta})  \\
   -\alpha e^{i\theta}(\partial_r+\frac{i}{r}\partial_{\theta}) & \lambda_{\mathbf{K}} + g/r  \end{array} \right), \\
 \mathcal{H}_{\mathbf{K'}}= \left( \begin{array}{cc}
 -\lambda_{\mathbf{K'}}+g/r & \alpha e^{i\theta}(\partial_r+\frac{i}{r}\partial_{\theta})  \\
   -\alpha e^{-i\theta}(\partial_r-\frac{i}{r}\partial_{\theta}) & \lambda_{\mathbf{K'}} +g/r \end{array} \right),
   \label{eq.hamiltonian}
\end{eqnarray}
where $\alpha$ is the Fermi velocity. 
Given the rotational symmetry of the impurity potential, we have expressed the Hamiltonian in polar coordinates. In the trivial case, the mass terms are equal, i.e., $\lambda_{\mathbf{K}} = \lambda_{\mathbf{K'}}=V$. In the non-trivial case, we have $\lambda_{\mathbf{K}} = -\lambda_{\mathbf{K'}} = 3\sqrt{3}t'$.  
We assume that the Coulomb potential does not couple the two valleys; as it decays as $1/q$ in momentum space, scattering with large momentum transfer is suppressed. 

\section{Bound states from a single valley}
The Coulomb impurity problem for a single valley is well studied\cite{Dong2003,Novikov2007} (see Appendix for explicit derivation). The valley Hamiltonians do not commute with angular momentum, $\hat{l}_z= i\hbar \partial_\theta$, but with $\hat{j}_z = \hat{l}_z +\frac{1}{2} \sigma_z$. The eigenstates are characterized by a quantum number $j = m + 1/2$, where $m$ is an integer. For a given $j$, there are several states indexed by an additional quantum number $\beta$ that takes non-positive integer values. The eigenstates in each valley take the form 
\begin{equation}
\Psi_{\mathbf{K},j,\beta} 
\!=\! \left(\! \begin{array}{c}
a_{j,\beta}(r)e^{im{\theta}}\\
b_{j,\beta}(r)e^{i(m+1){\theta}}\end{array} \!\right)\! ,
\Phi_{\mathbf{K'},j,\beta} 
\!=\! \left( \!\begin{array}{c}
c_{j,\beta}(r)e^{-im{\theta}}\\
d_{j,\beta}(r)e^{-i(m+1){\theta}}\end{array} \!\right)\!,
\label{eq.psiphi}
\end{equation}
where the radial wavefunctions $a_{j,\beta}(r)-d_{j,\beta}(r)$ are related to the radial part of the Hydrogen atom wavefunction. They are linear combinations of terms of the form $r^\chi e^{-r/a_0}$ ${}_1\mathcal{F}_1(r)$, where ${}_1\mathcal{F}_1(r)$ is a  confluent hypergeometric function. These represent impurity bound states with `Bohr radius' $a_0 \sim\alpha/\sqrt{\lambda_{\mathbf{K}/\mathbf{K'}}^2 - \epsilon^2}$, where $\epsilon$ is the energy eigenvalue which is itself determined by $j$, $\beta$, $\vert \lambda_{\mathbf{K}/\mathbf{K'}}\vert$ and $\alpha$. 

We note that bound states are formed even for infinitesimal $g$. For large $g$, eigenstates become unstable beyond a `supercritical' threshold, in analogy with the Hydrogen atom at $Z=137$\cite{landau1982,Pereira2007,Gupta2009,Gupta2008,Gamayun2011Ukr}. Beyond this instability, well-defined eigenstates may be obtained by regularizing the potential\cite{Pereira2008}. 
Indeed, our tight binding spectrum is well-defined for any $g$ due to implicit regularization at the scale of the lattice. As we are mainly concerned with chiral currents in Chern insulators, we always work in the subcritical regime in the interest of simplicity. 

\section{Inter-valley mixing}
At low energies, the Hamiltonian separates into the two valley-Hamiltonians, $\mathcal{H}_{\mathbf{K}}$ and $\mathcal{H}_{\mathbf{K'}}$. 
An arbitrary eigenstate may have components from both valleys, 
\begin{equation}
\psi(\mathbf{r}) \sim \mathcal{A} \Psi_{\mathbf{K},j,\beta} e^{i\bf{K}.\bf{r}} + \mathcal{B}  \Phi_{\mathbf{K'},j',\beta'} e^{-i\bf{K}.\bf{r}}, 
\label{eq.linearcombination}
\end{equation}
where $\Psi_{\mathbf{K},j,\beta}$ and $\Phi_{\mathbf{K'},j',\beta'}$ are eigenstates of $\mathcal{H}_{\mathbf{K}}$ and $\mathcal{H}_{\mathbf{K'}}$ respectively.
The exponential factors $e^{i\bf{K}.\bf{r}}$ and $e^{-i\bf{K}.\bf{r}}$ arise as each valley has average momentum $\mathbf{K}$ or $-\mathbf{K}$. 
 
\begin{figure}
  \includegraphics[width=\columnwidth]{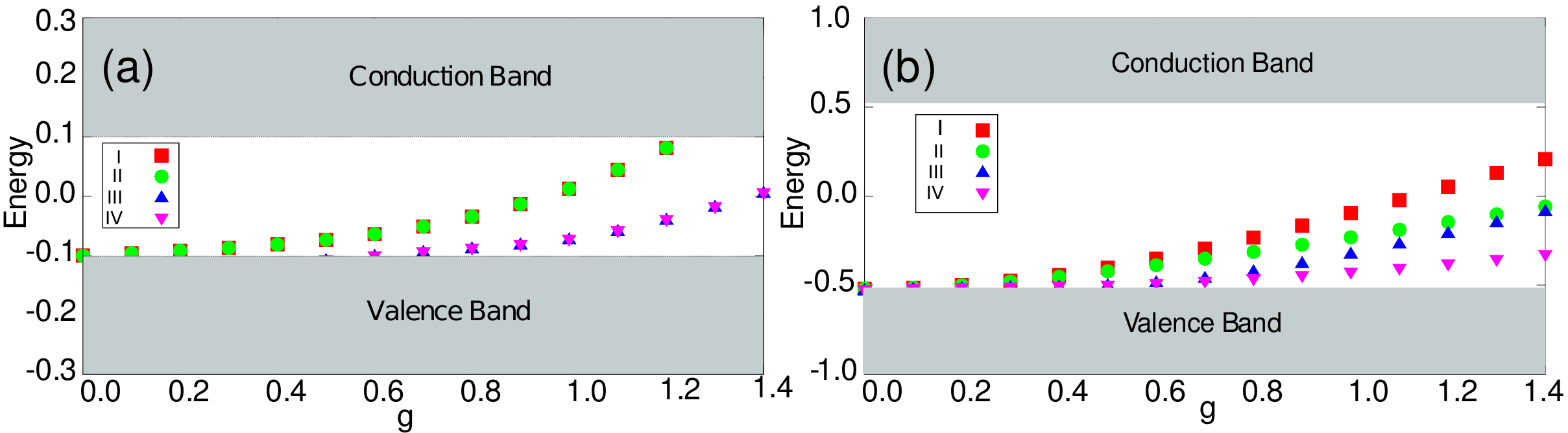}
 \caption{Bound state spectrum as a function of impurity strength for (a) trivial and (b) non-trivial gaps. We only show the four highest bound states. The shaded regions represent conduction and valence bands. }
  \label{fig.bs_spectrum}
 \end{figure}

In the trivial case, both valleys have the same mass term. Consequently, $\Psi_{\mathbf{K},j,\beta} e^{i\bf{K}.\bf{r}}$ and $\Phi_{\mathbf{K'},j,\beta} e^{-i\bf{K}.\bf{r}}$ are time-reversed pairs which are degenerate.
There is a weak inter-valley coupling arising from the Coulomb impurity which we have ignored in our analytic formulation. However, this is always present in our tight binding simulations.
Apparently, this inter-valley coupling does not mix states from the two valleys with the eigenstates remaining doubly degenerate\cite{Pereira2008}. This can be seen in the tight binding spectrum in Fig.~\ref{fig.bs_spectrum}(a). This degeneracy, in fact, arises from an approximate anti-unitary symmetry of the Hamiltonian 
that allows for a Kramers-like degeneracy (see Appendix). 
In contrast, in the non-trivial case shown in Fig.~\ref{fig.bs_spectrum}(b), the eigenvalues are split by the inter-valley coupling and are no longer degenerate.

 \section{Comparison with tight binding results}
We proceed to compare the analytic wavefunctions with tight binding results in the non-trivial case where bound states are non-degenerate. The analytically obtained wavefunction in Eq.~\ref{eq.linearcombination} is characterized by several parameters: $\{j,\beta\}$, $\{j', \beta' \}$ and the coefficients $\mathcal{A}$ and $\mathcal{B}$. Each bound state obtained from the tight binding numerics must correspond to an analytic solution with a certain choice of these parameters. 
In general, given a numerically obtained bound state, it is difficult to find the corresponding analytic parameters ($j,\beta,j',\beta',\mathcal{A},\mathcal{B}$) as it involves fitting many variables. 
However, for the highest energy bound state, this task is easier -- we take $\{j=1/2,\beta=0\}$ and $\{ j'=1/2,\beta' = 0 \}$ which correspond to the highest energy single-valley solutions. Moreover, as the single-valley solutions have the same energy, we assume a valley-symmetric linear combination with $\mathcal{A}=\mathcal{B}$. With these parameters, we obtain good agreement with the numerically obtained highest bound state.

To compare the analytic wavefunction with tight binding results, we compute the probability density. 
Using $\rho=\psi^{\dagger}\psi$,we find (see Appendix)  
\begin{equation}
\rho(r,\theta)\!=\! F^2(r) \! \left[ 4\lambda_{\mathbf{K}} \!+\! \frac{2\alpha}{a_0}\left\{\cos{(2\mathbf{K}\cdot \mathbf{r})} \!+\!\cos{(2\mathbf{K}\cdot \mathbf{r}-2\theta)}\right\}\right]\!,\!
\label{eq.density}
\end{equation}
where $F(r)=  
e^{-\frac{ r}{a_0}}r^{\eta-\frac{1}{2}}\mathcal{N}$, with $\eta = \sqrt{\frac{1}{4}-\frac{g^2}{\alpha^2}}$ and
$\mathcal{N}$ being a normalisation constant.

Fig.~\ref{fig:bound}(a,c,e) show the tight binding results (blue dots) for $\rho$ vs. distance from impurity, $r$. Comparing this with the analytical result presents two difficulties: (a) we do not have a closed form expression for the constant of normalization $\mathcal{N}$, and (b) the analytic wavefunction is `coarse-grained', i.e., it gives us a continuous function for density. To connect to the numerical result, we must integrate this result over an effective area corresponding to each lattice point. Once again, this does not give us a closed analytic form. To account for these two factors, we introduce a single fitting parameter -- an overall multiplicative constant. 
We then obtain excellent agreement between the analytic result and the tight binding calculations as seen in Fig.~\ref{fig:bound}(a,c,e). 

\section{Current in the bound state}

  \begin{figure*}
\includegraphics[width=6.75in]{{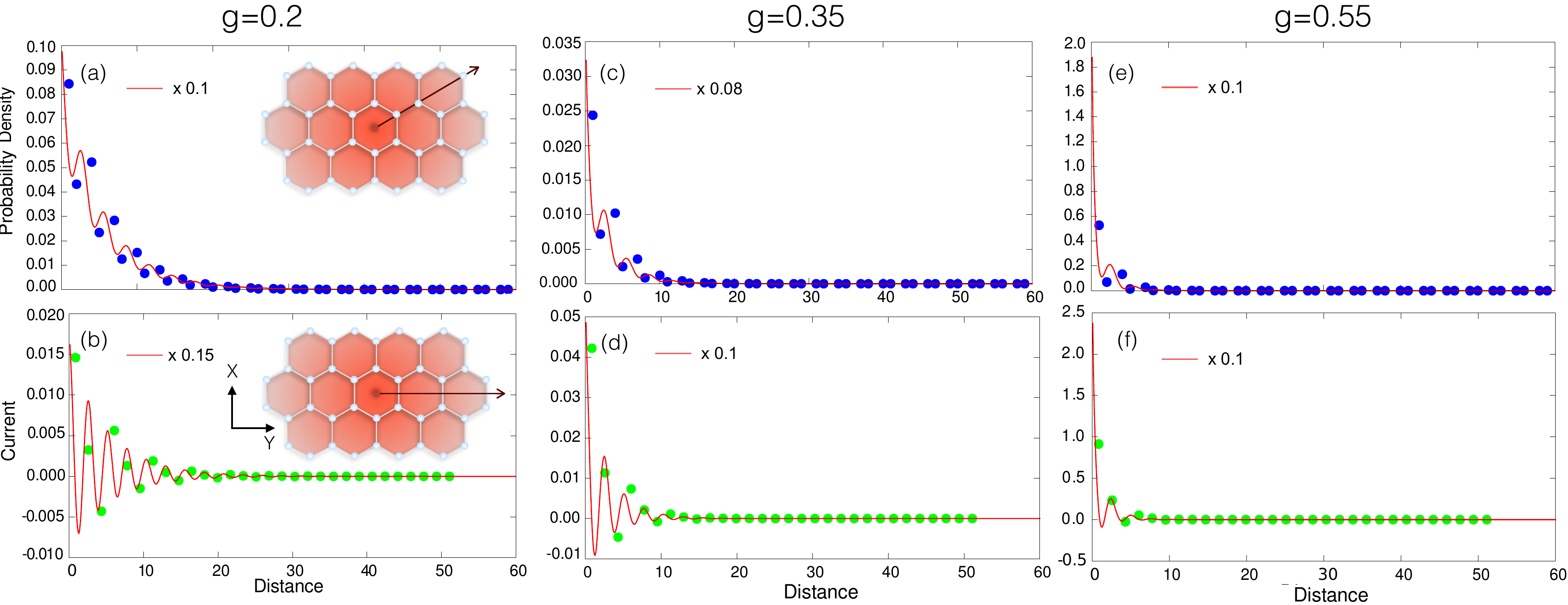}} 
  \caption{ Wavefunction of highest energy bound state for $g=0.2$ (a,b), $0.35$ (c, d) and $0.55$ (e,f). 
The upper panels plot probability density vs. distance from impurity at points along the line shown in the inset to (a). The lower panels show current vs. distance. Current is evaluated on bonds intersecting the line indicated in the inset to (b).     
Red lines show analytical results scaled by an overall factor as indicated in each panel. Dots are tight binding results.}
  \label{fig:bound}
  \end{figure*}
 
In the trivial case, bound states cannot carry currents due to time reversal symmetry.   
In the non-trivial case, we find that bound states carry a chiral current encircling the impurity.  
This is shown in Fig.~\ref{fig:bound}(b,d,f) with analytical and numerical results for the highest bound state. 
As with the probability density, we evaluate the current using the wavefunction of Eq.~\ref{eq.linearcombination} with the parameters $\{j=1/2,\beta=0\}$, $\{ j'=1/2,\beta' = 0 \}$ and $\mathcal{A}=\mathcal{B}$.
We find (see Appendix for details),
\begin{equation}
{\mathcal{J}}_{i,A}^{j,B}\approx 4 t F^2(R)\left( \mu \sin\{ \theta + \mathbf{K}\cdot\mathbf{\delta}\}+\lambda_\mathbf{K} \sin\{\theta- \mathbf{K}\cdot R\}\right)
\label{eq.J_boundstate}
\end{equation}
 where the bond centre has coordinates $\{R,\theta\}$ and $\mu =  \sqrt{\lambda_{\mathbf{K}}^2-\epsilon^2}$. 
As with the probability density, we scale this quantity by one multiplicative fitting parameter and find good agreement with numerics. 

\begin{figure}
 \includegraphics[width=\columnwidth]{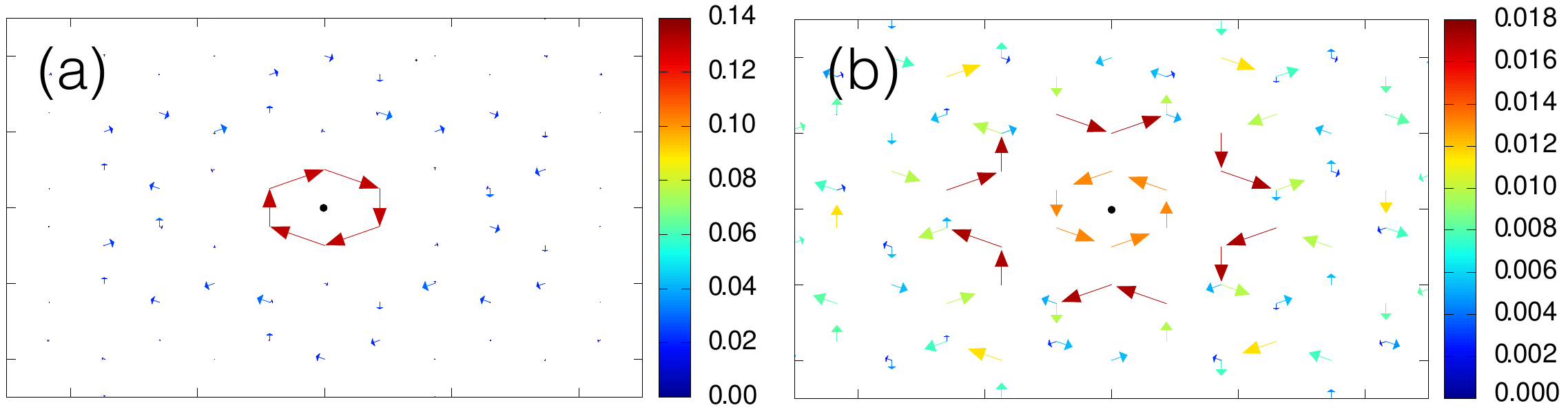}
 \caption {Tight binding results for current carried by bound states. (a) Current in the highest bound state for g = 0.72. (b) Current in the second highest bound state for g = 0.3. The color and length of an arrow show the amplitude of current on the bond; current direction is indicated by the arrow head. The color bar gives the current in units of $t$, the nearest neighbour hopping.}
  \label{fig.bs_current}
  \end{figure}

As shown in Fig.~\ref{fig.bs_current}, the bound state current circulates along loops surrounding the impurity. For simplicity, we show the current on nearest-neighbour bonds here, although next-nearest neighbour bonds also carry chiral currents. Typically, we find that the highest bound state has strongest current along the hexagon immediately surrounding the impurity, see Fig.~\ref{fig.bs_current}(a). In lower bound states, the current strength may peak further away from the impurity. The current direction may also change with increasing loop radius, see Fig.~\ref{fig.bs_current}(b). 

\section{Total current from linear response}
\label{sec.linearresponse}
\begin{figure}
\includegraphics[width=3.3in]{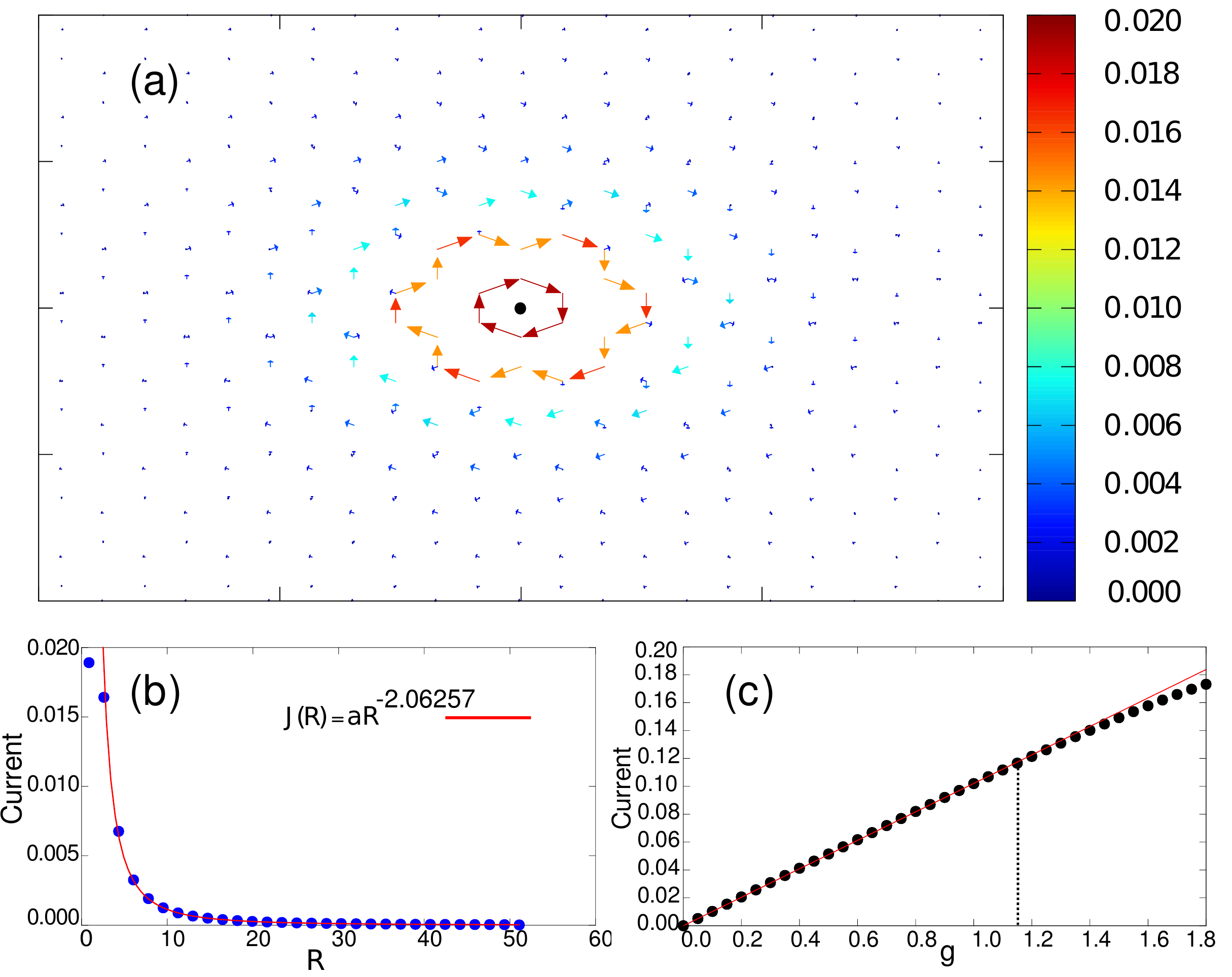}
  \caption{Tight binding results for total current around impurity.
  (a) Currents on bonds near impurity (g=0.5); the sites surrounding the impurity organize themselves into loops.
  (b) Current vs. distance from impurity for g = 0.5. The red line is the best fit (excluding the first two data points) to the form $J=aR^{-b}$. We find $b=2.06257 \pm 0.02286$, close to the linear response result of $b=2$.
  (c) Total current vs. $g$. The current is linear upto g $\sim$ 1.15. }
  \label{fig.linear_resp_current}
  \end{figure}

 Apart from individual bound states, tight binding calculations at half-filling show a total chiral current encircling the impurity in the non-trivial case, as shown in Fig.~\ref{fig.linear_resp_current}(a). 
 We describe this using a linear response approach below. 
 
In the clean Haldane model, the Green's function is 
\begin{equation}\label{eq:Green}
\mathcal{G}_{\gamma,\delta}^{0} (\mathbf{k},\omega)= \langle   \bar{\psi}_{\mathbf{k,\omega},\gamma} {\psi}_{\mathbf{q},\omega,\delta}\rangle = \sum_{l = \pm 1} \frac{\mathcal {T}_{\gamma l} \mathcal {T}{^\dagger}_{l \delta}}{i\omega - \epsilon_{k,l}}.
\end{equation}
Here, $\psi$'s are Grassmann numbers, $\gamma,\delta=A/B$ are sublattice indices, and $\omega=(2n+1)\pi k_B T$ are fermionic Matsubara frequencies. 
The matrix $\mathcal{T}$ diagonalizes the $2\times2$ Hamiltonian matrix, giving eigenvalues $\epsilon_{k,l}$, with $l$ being the band index. We have   
\begin{equation}\label{eq:diagonal}
\mathcal{T}=\left( \begin{array}{cc}
\cos(\frac{\theta_\mathbf{k}}{2}) & -e^{-i\phi_\mathbf{k}}\sin(\frac{\theta_\mathbf{k}}{2})\\
e^{i\phi_\mathbf{k}}\sin(\frac{\theta_\mathbf{k}}{2}) & \cos(\frac{\theta_\mathbf{k}}{2})
\end{array} \right). 
\end{equation}
Near the Dirac points at $\pm\mathbf{K}$, the 
parameters $\theta_\mathbf{k}$ and $\phi_\mathbf{k}$
take simple forms, given by  
$ \cos(\theta_\mathbf{k}) = \frac{\mp \lambda_{\mathbf{K}}}{\sqrt{k_{x}^2 + k_{y}^2 + \lambda_{\mathbf{K}}^2}}$ and $e^{i\phi_\mathbf{k}} = \frac{\pm k_{y} - i k_{x}}{\sqrt{k_{x}^2 + k_{y}^2}}$. We have set the chemical potential to zero as we are at half-filling, and the fermi velocity to unity for simplicity.

We treat the impurity potential as a perturbation. 
Taking the impurity potential to be small, $g\ll \lambda_{\mathbf{K}}, \lambda_{\mathbf{K'}}$, the perturbing term in the action is given by  
\begin{equation}
S_1 \!=\!
\frac{2 \pi g}{N}\sum_{\mathbf{k},\mathbf{q},\omega}\frac{1}{q}\left[ \bar{\psi}_{\mathbf{k},\omega,A} {\psi}_{\mathbf{k}+\mathbf{q},\omega,A} +  \bar{\psi}_{\mathbf{k},\omega,B} {\psi}_{\mathbf{k}+\mathbf{q},\omega,B}\right]\!,  
\end{equation}
obtained by Fourier transforming the Coulomb impurity potential.
We are interested in the current on a given bond connecting $(i,A)\rightarrow (j,B)$, given by the operator
\begin{equation}
 \mathcal{J}_{i,A}^{j,B}  = i t(\bar{\psi}_{i,A,\tau}\psi_{j,B,\tau}-\bar{\psi}_{j,B,\tau}\psi_{i,A,\tau}).
\end{equation}
 Using standard linear response theory, 
\begin{equation}
\langle \mathcal{J}_{i,A}^{j,B} \rangle -  \langle \mathcal{J}_{i,A}^{j,B} \rangle_0 =  \langle \mathcal{J}_{i,A}^{j,B} \rangle_0  \langle S_1 \rangle_0 - \langle S_1 \mathcal{J}_{i,A}^{j,B} \rangle_0.
\end{equation}
The notation $\langle \{.\}\rangle_0$ denotes expectation value with respect to the unperturbed action. The second term on the left hand side vanishes as current is zero for inter-sublattice bonds in the clean limit. The right hand side can be evaluated using Wick's theorem to give
\begin{eqnarray}
 \nonumber
 \langle \mathcal{J}_{i,A}^{j,B} \rangle &=&\frac{i t}{N^2\beta} \sum_{{\bf k},{\bf q},\omega,\kappa}\frac{2\pi g}{{q}}\left[ \mathcal{G}^0_{B,\kappa}({\bf k},\omega)\mathcal{G}^0_{\kappa, A}({\bf k}+{\bf q},\omega)e^{-i\theta_1}\right. \\ &-&
\left.  \mathcal{G}^0_{A, \kappa}({\bf k},\omega)\mathcal{G}^0_{\kappa, B}({\bf k}+{\bf q},\omega)e^{-i\theta_2}\right],
 \end{eqnarray}
where, $\theta_1= {\bf k}\cdot{\bf r_j}-({\bf k}+{\bf q})\cdot {\bf r_i}$ and $\theta_2 = {\bf k}\cdot{\bf r_i}-({\bf k}+{\bf q})\cdot {\bf r_j}$.  The index $\kappa$ sums over sublattices A and B, while N is the number of sites in the system. We obtain
 \begin{eqnarray}\label{eq:freq}
 \langle \mathcal{J}_{i,A}^{j,B} \rangle = k_B T \sum_{{\bf k},{\bf q},\omega,\kappa,\l,\l'}\frac{F_{l l',\kappa}}{(i \omega - \epsilon_{{\bf k},l})(i \omega - \epsilon_{{\bf k+q},l'})},
 \label{eq:current}
 \end{eqnarray}
 where 
 $F_{ll', \kappa}$ is given by
\begin{eqnarray}
\nonumber  F_{ll', \kappa}\!\!&=& \!\! i t \frac{2 \pi g}{N^2  q}\left[ \mathcal T_{\kappa l}({\bf k} + {\bf q})\mathcal T_{l A}^\dagger ({\bf k} + {\bf q})\mathcal T_{B l'}({\bf k})\mathcal T_{l' \kappa}^\dagger ({\bf k}) e^{-i\theta_1} \right.\\ 
&-& \!\! \left.
 \mathcal T_{A l}({\bf k})\mathcal T_{l \kappa}^\dagger ({\bf k})\mathcal T_{\kappa l'}({\bf k}+{\bf q})\mathcal T_{l' B}^\dagger ({\bf k}+{\bf q})e^{-i\theta_2}\right].
\label{eq:matsubara}
\end{eqnarray}
In the momentum sums in Eq.~\ref{eq:current}, we only keep intra-valley terms, i.e., $\mathbf{k}$ and $\mathbf{k}+\mathbf{q}$ are taken to be in the same valley. This is justified due to the factor of $\frac{1}{q}$ in Eq.~\ref{eq:matsubara}, which originates from the Fourier transform of the Coulomb potential. This strongly suppresses inter-valley terms which require a large ${q}$.

To better understand the current distribution, we denote a given bond as $(\mathbf{R},\mathbf{\delta})$, where $\mathbf{R}$ is the distance from the impurity to the centre of the bond and $\mathbf{\delta}$ is the bond vector. After summing over Matsubara frequencies, assuming zero temperature, 
we obtain the bond current 
\begin{eqnarray}
\nonumber \langle \mathcal{J}_{i,A}^{j,B} \rangle=  \frac{t}{N^2} \sum_{\mathbf{k},\mathbf{q}} \frac{ 4 \pi g \lambda_{\mathbf{K}} \cos({\bf k \cdot \delta}) \cos(\frac{{\bf q}}{2}\cdot \delta) }{ q (-\zeta_{\mathbf{k}} - \zeta_{\mathbf{k}+\mathbf{q}})\zeta_{\mathbf{k}+\mathbf{q}} \zeta_{\mathbf{k}}} \times
\\
 \left\{q_x \sin({\bf K \cdot \delta)} + q_y \cos({\bf K \cdot \delta)}\right\} \sin({\bf q}\cdot{\bf R})
 ,
\label{eq:current3}
\end{eqnarray}
where $\zeta_{\mathbf{k}}  =  \sqrt{k_x^2+k_y^2+\lambda_{\mathbf{K}}^2}$ is the absolute value of the excitation energies at $\mathbf{k}$.
We notice that the current depends on $\mathbf{R}$ solely through the $\sin({\bf q}\cdot{\bf R})$ term above. We are interested in the current distribution far from the impurity, when $\vert \mathbf{R}\vert \gg a $ (the lattice spacing). As the $\sin({\bf q}\cdot{\bf R})$ term is highly oscillatory, the dominant contribution in the above summation will come from small $\mathbf{q}$ values. Taking $\bf R$ along y and $\bf \delta$ along x direction (see Fig.~\ref{fig:bound}(b)), we rewrite the summation as an integral,
\begin{equation}\label{eq:current4}
\langle \mathcal{J}_{i,A}^{j,B} \rangle \propto \int dq_y q_y  \sin (q_{y} R).
\end{equation}
We have kept only the leading order term in $q_y$ as the summation is dominated by small $q_y$ values. 
The integral scales as $q_{y}^2$. As $R$ is the only distance scale, the bond current will scale as $1/R^2$ for large R. Indeed, this is confirmed by our tight binding analysis. Fig.~\ref{fig.linear_resp_current}(b)  shows the current as a function of $R$ for $g = 0.5$. For large $R$ (excluding two bonds closest to the impurity), we obtain a good fit to the form $J=aR^{-b}$ with $b\approx2$. 
 Thus, remarkably, the total bond current induced by the impurity decays algebraically as $R^{-2}$ in contrast to the exponentially decaying bound state currents. We surmise that the dominant contribution to currents comes from extended states of the valence band, rather than from bound states.
 
The above linear response calculation is justified for small $g$. However, in our numerical results, the total current scales linearly with $g$ over a large window, as shown in Fig.~\ref{fig.linear_resp_current}(c). Here, total current is defined as the sum of bond currents on all bonds intersecting the indicated line in Fig.~\ref{fig:bound}(b). Surprisingly, deviation from linearity sets in at $g\sim 1.15$ which is the `supercritical' threshold at which the energy of the highest bound state becomes zero\cite{Pereira2008}, see Fig.~\ref{fig.bs_spectrum}(b).   

\section{Coexisting trivial and non-trivial masses}
We have focussed on two particular cases above -- the trivial ($\lambda_{\mathbf{K}} = \lambda_{\mathbf{K}'}$) and the non-trivial ($\lambda_{\mathbf{K}} = -\lambda_{\mathbf{K}'}$). However, these are two limiting cases of the general Hamiltonian with both mass terms, i.e., $\lambda_{\mathbf{K}} \neq \pm\lambda_{\mathbf{K}'}$. In the tight binding simulations, this general Hamiltonian is realized by keeping both the sublattice potential and the complex $t'$ hopping. The topological character of the phase is determined by the relative signs of the mass terms: a trivial (topological) phase is realized when $\lambda_{\mathbf{K}}$ and $\lambda_{\mathbf{K}'}$ have the same (opposite) sign\cite{Haldane1988,Gonzalez2012}. Indeed, these phases are separated by a topological phase transition which occurs when the net mass term in one valley vanishes leading to a closing of the electronic gap.  

Our tight binding simulations reveal that the total chiral current induced by the impurity bears a signature of this topological phase transition. On the trivial side, i.e., with $\mathrm{sign}(\lambda_{\mathbf{K}}) = \mathrm{sign}(\lambda_{\mathbf{K}'})$,
we find that the current is well described by an exponentially decaying function. This is in agreement with conventional wisdom about the impurity response of an insulator with the band gap setting a length scale for any impurity response. In contrast, on the topological side, the current decays algebraically with a $1/R^2$ dependence. In Sec.~\ref{sec.linearresponse} above, this algebraic dependence is justified using a linear response approach in the Haldane limit ($\lambda_{\mathbf{K}} = -\lambda_{\mathbf{K}'}$). Extending the linear response argument to the general case with $\lambda_{\mathbf{K}}  \neq \pm \lambda_{\mathbf{K}'}  $, we find that the current expression is given by,
\begin{eqnarray}
\nonumber \langle \mathcal{J}_{i,A}^{j,B} \rangle \propto \int d^2 k dq_x  dq_y \bigg[  \alpha q_y  \sin (q_{y} R) + \beta  \sin (q_{y} R) \bigg] \times \\
 \frac{f(t', V, \vert \mathbf{q} \vert, \vert \mathbf{k} + \mathbf{q} \vert)}{\vert \mathbf{q}\vert},\phantom{ab}
\label{eq.J_integral}
\end{eqnarray}
where $f(t', V, \vert \mathbf{q} \vert, \vert \mathbf{k} + \mathbf{q} \vert)$ is a complicated function, whose precise nature is not important for extracting the long distance behaviour. 
In the non-trivial limit, we only have the first term within the square brackets. In the trivial limit, we only have the second. Of course, in the trivial limit, time reversal symmetry forces the current to be strictly zero.

To deduce the long distance behaviour of the current, we note that the dependence on $\mathbf{R}$ arises solely from the $\sin (q_{y} R)$ term. Far away from the impurity, this term is highly oscillatory. We expect the dominant contribution to arise from very small $q_y$. Here, we assume that the dominant contribution comes from $q_x \neq 0$, in which case, the $1/\vert \mathbf{q} \vert$ terms goes to $q_x^{-1}$ in the small $q_y$ limit. In addition, the function $f(t', V, \vert \mathbf{q} \vert, \vert \mathbf{k} + \mathbf{q} \vert)$ has a finite $q_y\rightarrow 0$ limit. 
By this reasoning, we expect that the first term scales as $\int dq_y q_y \sin(q_y R)$, which leads to the $1/R^2$ dependence for the current as argued in Sec.~\ref{sec.linearresponse}. 

Similary, the second term is dominated by its behaviour in the small $q_y$ limit. Na\"ively, we can argue as before that this integral will scale as $\int dq_y \sin(q_y R)$ leading to a $1/R$ dependence. However, when $q_x$ is itself small, this integral will scale as $\int dq_y q_y^{-1}\sin(q_y R)$ which is dimensionless. In this case, it is plausible that the result is an exponentially decaying current $J\sim C\exp(-D/R)$. Our tight binding results support this picture. 

\begin{figure}
\includegraphics[width=3.25in]{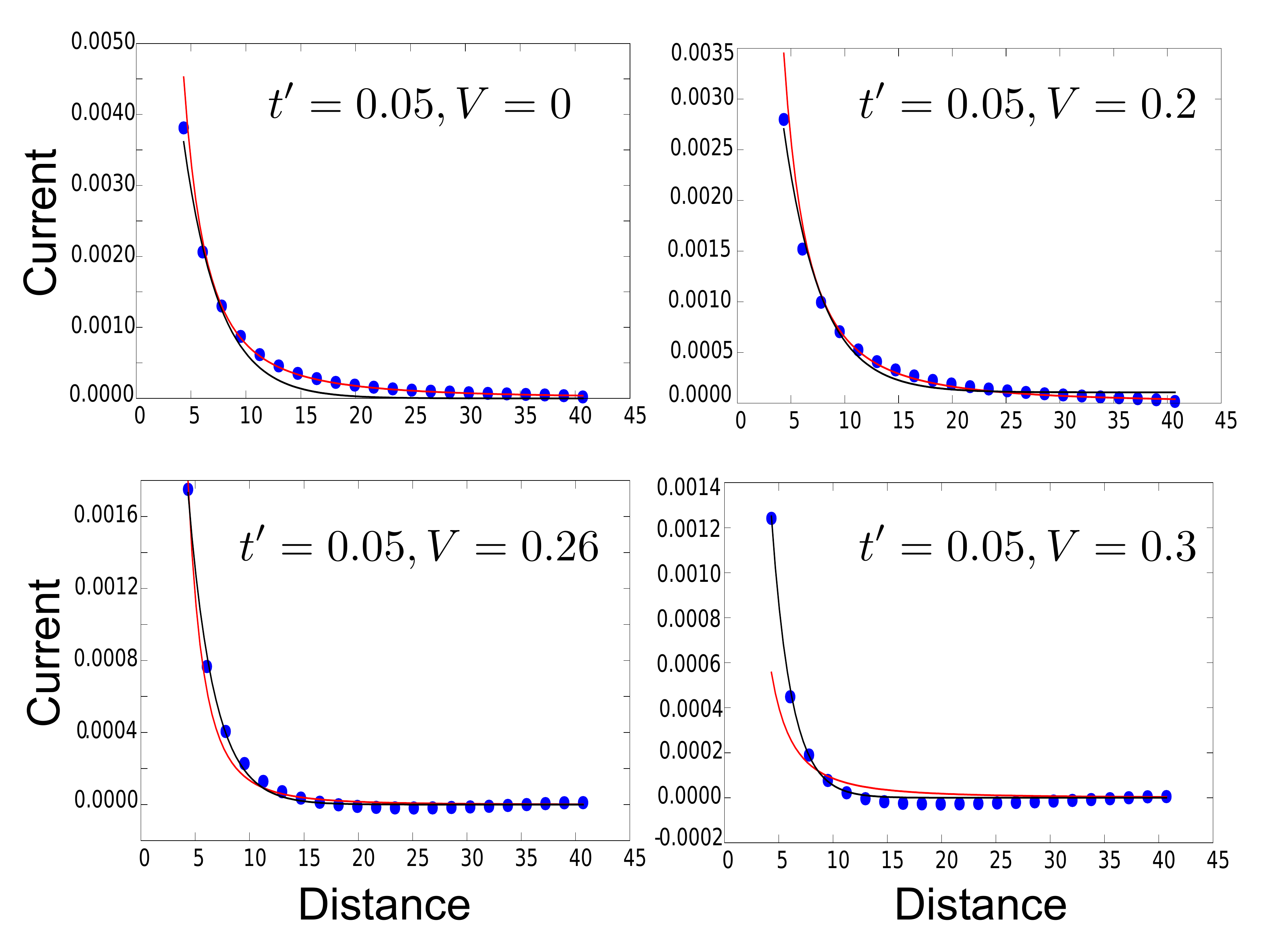}  
\caption{Current vs. $R$ for the general Hamiltonian with both mass terms, with $g=0.3$. Keeping $t'$ fixed, we increase the sublattice potential $V$. A topological transition occurs at the $V\sim 0.2598$. In each case, we plot the tight binding results for current with the best fit to an exponential function $J = Ae^{-B R}$ (black line) and to a polynomial function $J = CR^{-D}$ (red line). (a \& b) With both mass terms present, but staying on the topological side, the polynomial function gives a better fit. The exponent is $D\approx2$. (c \& d) On the trivial side, the exponential function gives a better fit.  }
\label{fig.top_transition}
\end{figure}

The current is thus a sum of power-law and exponential terms. We cannot find the relative strengths of these terms analytically, as the integrals do not have a closed form solution. Numerically, we find that exponential decay dominates in the trivial phase ($\mathrm{sign}(\lambda_{\mathbf{K}}) = \mathrm{sign}(\lambda_{\mathbf{K}'})$) while the power law term dominates in the topological phase ($\mathrm{sign}(\lambda_{\mathbf{K}}) \neq \mathrm{sign}(\lambda_{\mathbf{K}'})$). 
This is shown in Fig.~\ref{fig.top_transition} where we fix $t'=0.05$ and tune the sublattice potential $V$. This is equivalent to tuning the relative strengths of the mass terms. A trivial-topological phase transition, accompanied by closing of the electronic band gap, occurs at $V=0.2598$. For different values of $V$, we fit the tight binding results for current to an exponential as well as to a power-law. On the topological side of the transition, we find excellent agreement with a polynomial fit; the resulting power-law exponent is close to ${-2}$. On the trivial side, it is the exponential fit that better describes the data. 

In summary, the breaking of time-reversal symmetry leads to chiral currents around the impurity. However, the precise distribution of currents bears a signature of the topological character of the phase.

\section{Discussion}
We have shown that impurities in Chern insulators induce local chiral currents, as a consequence of time-reversal symmetry breaking.  
Specifically, we have studied the response to a Coulomb impurity in two gapped Dirac systems: (a) the trivial insulator with equal mass terms ($\lambda_{\mathbf{K}} = \lambda_{\mathbf{K}'}$), and (b) the non-trivial Haldane insulator with opposite mass terms $(\lambda_{\mathbf{K}} = -\lambda_{\mathbf{K}'})$. While the former does not develop any currents, the latter develops current-carrying bound states and a net chiral current that decays as $1/R^2$. 
These two limiting cases allow for an analytic calculation of the current. 
However, both types of masses can coexist giving rise to a topological transition realized by tuning their relative strengths. This transition is reflected in the total chiral current encircling the impurity -- in the non-trivial phase, the chiral current decays as $1/R^2$ while in the trivial phase, it decays exponentially. Thus, although the existence of a chiral current is only a signature of time-reversal breaking, the distribution of the current bears a signature of topology.

Our results could be tested with the recent cold atom realization of the Haldane model\cite{Jotzu2014} and with graphene-like candidate materials\cite{Mao2016}. 
An impurity-induced current should give rise to localized magnetic fields that can be measured using scanning-probe magnetometry\cite{Taylor2008}. 
Our results also apply to weak topological insulators which can be thought of as two copies of a Chern insulator, one for each spin\cite{KaneMele2005}. An impurity will induce currents of opposite directions in each spin sector leading to a chiral spin current rather than a charge current. Similar features may occur in other topological phases -- particularly in chiral superconductors for which several candidates have been proposed, e.g.,  SrRu$_2$O$_4$\cite{Kallin2009}, doped graphene\cite{Pathak2010,Nandkishore2012,BlackSchaffer2012} and TiSe$_2$\cite{Ganesh_2014}.  
 
\acknowledgments{We thank Arnab Sen and Kumar S. Gupta for discussions. We thank Igor Herbut for the arguments in Appendix A3.} 
 
\renewcommand{\theequation}{A\arabic{equation}}
\renewcommand{\thefigure}{A\arabic{figure}}
\setcounter{equation}{0}
\setcounter{figure}{0}
\appendix
\section{Single valley solutions}
We follow the derivation of bound state wavefunctions presented in Refs.~\onlinecite{Novikov2007} and \onlinecite{Dong2003}. The discussion in these references is at the single-valley level; we emphasize the differences between the trivial and non-trivial cases here.

\subsection{Non-Trivial case}
 
The Hamiltonians discussed in the main text (Eq.~\ref{eq.hamiltonian}) have spinor solutions of the form given in Eq.~\ref{eq.psiphi}. We discuss this for the $\mathbf{K}$ valley below; similar arguments apply for the $\mathbf{K'}$ valley as well. Performing a transformation of the form $a(r)\sqrt{r}=A(r)$ and $b(r)\sqrt{r}=B(r)$, we obtain the differential equations:
\begin{eqnarray}
\nonumber B^{\prime}(r)+\frac{j}{r}B(r)=\frac{1}{\alpha}(\epsilon+\lambda_{\mathbf{K}}-\frac{g}{r})A(r), \\
 A^{\prime}(r)-\frac{j}{r}A(r)=-\frac{1}{\alpha}(\epsilon-\lambda_{\mathbf{K}}-\frac{g}{r})B(r).
 \label{eq.Aprime1}
 \end{eqnarray}
We are interested in bound states in the gap, $-\lambda_{\mathbf{K}}<\epsilon<\lambda_{\mathbf{K}}$. A coordinate scaling of the form $\rho=2\mu r$, with $\mu = \sqrt{\lambda_{\mathbf{K}}^2-\epsilon^2}$, gives 
 \begin{eqnarray}
\nonumber B^{\prime}(\rho)+\frac{j}{\rho}B(\rho)=\left(-\frac{g}{\alpha\rho}+\frac{1}{2\alpha}\sqrt{\frac{\lambda_{\mathbf{K}}+\epsilon}{\lambda_{\mathbf{K}}-\epsilon}}\right)A(\rho),\\
A^{\prime}(\rho)-\frac{j}{\rho}A(\rho)=\left(\frac{g}{\alpha\rho}+\frac{1}{2\alpha}\sqrt{\frac{\lambda_{\mathbf{K}}-\epsilon}{\lambda_{\mathbf{K}}+\epsilon}}\right)B(\rho).
\label{eq.Aprime2}
\end{eqnarray}
For notational convenience, we denote $\mu_{\pm}=\sqrt{\lambda_{\mathbf{K}}\pm\epsilon}$.
The above differential equations are coupled and can be separated\cite{Novikov2007} by taking $B(\rho)=\mu_+(f_{1,{\mathbf{K}}}-f_{2,\mathbf{K}})$ and $A(\rho)=\mu_-(f_{1,\mathbf{K}}+f_{2,\mathbf{K}})$.  Substitution in Eqs.~\ref{eq.Aprime2} gives
\begin{eqnarray}
\nonumber f_{2,\mathbf{K}}^{\prime}+f_{2,\mathbf{K}}(\frac{1}{2\alpha}+\frac{\tau_2}{\alpha\rho})=f_{1,\mathbf{K}}(\frac{j}{\rho}+\frac{\tau_1}{\alpha\rho}), \\
f_{1,\mathbf{K}}^{\prime}-f_{1,\mathbf{K}}(\frac{1}{2\alpha}+\frac{\tau_2}{\alpha\rho})=f_{2,\mathbf{K}}(\frac{j}{\rho}-\frac{\tau_1}{\alpha\rho}),
\label{eq.f1prime1}
\end{eqnarray}
where $\tau_1=\frac{g\lambda}{\mu}$ and $\tau_2=\frac{g\epsilon}{\mu}$.
These equations can be decoupled to resemble the general Tricomi equation\cite{Dong2003},
\begin{equation}
f_{1/2,\mathbf{K}}^{\prime\prime} +\frac{1}{\rho} f_{1/2,\mathbf{K}}^{\prime}
+\left(-\frac{1}{4\alpha^2}-\frac{\tau_2\pm\frac{\alpha}{2}}{\alpha^2\rho}-\frac{\eta^2}{\rho^2}\right)f_{1/2,\mathbf{K}}=0,
\label{eq.diffeqn1,2}
\end{equation}
where, $\eta=\sqrt{j^2-g^2/\alpha^2}$. The connection to the Hydrogen atom is now readily seen -- the above equation is identical to the radial equation of the Hydrogen atom problem.  
The solutions of  Eq.~\ref{eq.diffeqn1,2} for $f_{1,\mathbf{K}}(\rho)$ and $f_{2,\mathbf{K}}(\rho)$ are
\begin{eqnarray}
\nonumber f_{1,\mathbf{K}}(\rho)&=&{}_1\mathcal{F}_1(1+\frac{\tau_2}{\alpha}+\eta,1+2\eta,\frac{\rho}{\alpha})e^{-\frac{\rho}{2\alpha}+\eta\log{\rho}} c_1,
 \\
f_{2,\mathbf{K}}(\rho)&=&{}_1\mathcal{F}_1(\frac{\tau_2}{\alpha}+\eta,1+2\eta,\frac{\rho}{\alpha})e^{-\frac{\rho}{2\alpha}+\eta\log{\rho}} c_2,
\label{eq.f2analyticalK}
\end{eqnarray}
where ${}_1\mathcal{F}_1(a,c;z)=1+\frac{az}{(c) 1!}+\frac{a(a+1)z^2}{c(c+1)2!} + \ldots$ is the confluent hypergeometric function. The constants $c_1$ and $c_2$ are not independent; their ratio can be determined by examining Eqs.~\ref{eq.f1prime1} in the limit $\rho \rightarrow 0$. We have
\begin{equation}
\frac{c_1}{c_2} = \frac{\frac{\tau_2}{\alpha} + \eta}{ \frac{\tau_1}{\alpha} + j}.
\label{eq.cratios}
\end{equation}

When $a$ is a non-positive integer, the expansion of $_1\mathcal{F}_1(a,c;z)$ in powers of $z$ terminates at some order. $_1\mathcal{F}_1$ then reduces to a polynomial so that 
$f_{1/2,\mathbf{K}}$ decays to zero as $\rho\rightarrow \infty$ and is square-integrable\cite{landau1982}. 
We obtain the energy spectrum from this condition by writing $\frac{\tau_2}{\alpha}+\eta=\beta$, where $\beta$ is a non positive integer. We obtain the energy eigenvalue $\epsilon = {-\vert \lambda_\mathbf{K} \vert }/{\sqrt{1+ \frac{g^2}{\alpha^2 (\beta - \eta)^2}     } }$.

The wavefunction for the highest bound state, obtained by setting $\beta=0$, takes a somewhat simple form. The first argument of the confluent hypergeometric function in $f_{2,\mathbf{K}}(\rho)$ becomes zero; as a result, the confluent hypergeometric function in $f_{2,\mathbf{K}}(\rho)$ becomes unity. With $\beta = 0$, the confluent hypergeometric function in $f_{1,\mathbf{K}}(\rho)$ does not reduce to a polynomial -- however, this is preempted by the coefficient $c_1$ going to zero according to Eq.~\ref{eq.cratios}.

Similar arguments give us solutions for the $\mathbf{K'}$ valley, with $C(\rho)=\mu_+(f_{1,\mathbf{K'}}+f_{2,\mathbf{K'}})$ and $D(\rho)=\mu_-(f_{1,\mathbf{K'}}-f_{2,\mathbf{K'}})$, where $c(r)\sqrt{r}=C(r)$ and $d(r)\sqrt{r}=D(r)$.
This leads to 
\begin{eqnarray}
\nonumber
f_{1,\mathbf{K'}}(\rho)=_1\!\!\mathcal{F}_1(\frac{\tau_2}{\alpha}+\eta,1+2\eta,\frac{\rho}{\alpha})c_2e^{-\frac{\rho}{2\alpha}+\eta\log{\rho}}, \\
 f_{2,\mathbf{K'}}(\rho)=_1\!\!\mathcal{F}_1(1+\frac{\tau_2}{\alpha}+\eta,1+2\eta,\frac{\rho}{\alpha})c_1e^{-\frac{\rho}{2\alpha}+\eta\log{\rho}}.
\label{eq.f1kprimenontrv}
\end{eqnarray}
Comparing the solutions at the two valleys, we find that $f_{1,\mathbf{K}^{\prime}}
(\rho)=f_{2,\mathbf{K}}(\rho)$ and $f_{2,\mathbf{K}^{\prime}}
(\rho)=f_{1,\mathbf{K}}(\rho)$.

Retracing our sequence of transformations, we obtain the radial wavefunctions $a(r)-d(r)$ in spinors $\Psi_{\mathbf{K}}$ and $\Phi_{\mathbf{K'}}$ (Eq.~5 of main text),
\begin{eqnarray}
\nonumber a(r), c(r)&=& \mathcal{N}\mu_{\mp}e^{-\frac{\mu r}{\alpha}}r^{\eta-\frac{1}{2}} \times \\
\nonumber  &\phantom{a}&  \left({}_1\mathcal{F}_1(1+\frac{\tau_2}{\alpha}+\eta,1+2\eta;2\frac{\mu}{\alpha} r)c_1 \right.  + \\
&\phantom{a}& \left.  {}_1\mathcal{F}_1(\eta+\frac{\tau_2}{\alpha},1+2\eta;2\frac{\mu}{\alpha} r)c_2\right),
\end{eqnarray}
and
\begin{eqnarray}
\nonumber b(r), d(r)&=& \pm\mathcal{N}\mu_{\pm}e^{-\frac{\mu r}{\alpha}}r^{\eta-\frac{1}{2}} \times \\
&\phantom{a}& \nonumber \left({}_1\mathcal{F}_1(1+\frac{\tau_2}{\alpha}+\eta,1+2\eta;2\frac{\mu}{\alpha} r)c_1 \right. \\
&\phantom{a}& \left. -{}_1\mathcal{F}_1(\eta+\frac{\tau_2}{\alpha},1+2\eta;2\frac{\mu}{\alpha} r)c_2\right),
\label{eq.analyticalformofwavefunction}
\end{eqnarray}
where `$\pm$' denotes $+$ for the $\mathbf{K}$ valley and $-$ for the $\mathbf{K}^{\prime}$ valley respectively. We have introduced $\mathcal{N}$ as an overall normalization constant, to be determined later. 
The constants $c_1$ and $c_2$ are not independent, with their ratio fixed as before. It is easy to see that the radial wavefunctions from the two valleys are not time reversed pairs. If that were the case, the radial wavefunctions would have been identical, i.e., $a(r) = c(r)$ and $b(r) = d(r)$.

\subsection{Trivial case - solutions from each valley}
In the trivial case, the solutions again have the the spinor form of Eq.~5 of the main text. But, the radial part separates out such that the equations are the same for both valleys. It is sufficient to find analytical expressions for the functions $f_{1,\mathbf{K}}$ and $f_{2,\mathbf{K}}$ for the $\mathbf{K}$ valley. We denote $f_{1,\mathbf{K}/\mathbf{K'}} = f_1$ and $f_{2,\mathbf{K}/\mathbf{K'}} = f_2$. The radial wavefunctions take the form $A(\rho),C(\rho)=\mu_+(f_1+f_2)$ and $B(\rho),D(\rho)=\mu_-(f_1-f_2)$
Following the same procedure as in the non trivial case, we obtain the following differential equations,
 \begin{equation}
 f_{1,2}^{\prime\prime} +\frac{1}{\rho} f_{1,2}^\prime
 +\left( -\frac{1}{4\alpha^2}-\frac{\tau_2\mp\frac{\alpha}{2}}{\alpha^2\rho}-\frac{\eta^2}{\rho^2}\right)f_{1,2}=0.
 \label{eq:trivial}
 \end{equation}
This equation also has solutions containing confluent hypergeometric functions similar to Eqs.~\ref{eq.f1kprimenontrv}. It follows that the radial wavefunctions in both valleys are identical.
It is easy to see that the solutions from the two valleys, $\Psi_{\mathbf{K},j,\beta}$ and $\Phi_{\mathbf{K}',j,\beta}$, are time-reversed pairs; they are related by a straightforward complex conjugation operation.
 
 \subsection{Degeneracy in trivial case}
 In the trivial case, the full Hamiltonian can be written as a $4\times 4$ matrix, given by
\begin{equation}
H_{\mathbf{K},\mathbf{K}'} \sim \left(
\begin{array}{cccc}
g/r -\lambda & \hat{p}_y + i\hat{p}_x & \epsilon g/r & 0 \\
\hat{p}_y - i\hat{p}_x  & g/r +\lambda & 0 & \epsilon g/r \\
\epsilon g/r & 0 &  g/r -\lambda & -\hat{p}_y + i\hat{p}_x \\
0 & \epsilon g/r & -\hat{p}_y - i\hat{p}_x & g/r + \lambda
\end{array}
\right).
\end{equation}
We have taken $\lambda_{\mathbf{K}} = \lambda_{\mathbf{K}'} =\lambda$. We have the set the Fermi velocity to unity for simplicity. Operators $\hat{p}_{x/y}$ are momentum operators. 
The diagonal blocks of the Hamiltonian contain intra-valley terms, while the off-diagonal blocks encode weak inter-valley scattering from the Coulomb impurity potential. We have represented this term as $\epsilon g/r$, where $\epsilon$ is a small parameter proportional to $1/\vert \mathbf{K} \vert$. 
This Hamiltonian may be rewritten as 
\begin{eqnarray}
\nonumber H_{\mathbf{K},\mathbf{K}'} \sim  g/r \{\sigma_0 \otimes \sigma_0 \}- \lambda \{\sigma_z \otimes \sigma_0\} + \hat{p}_y \{\sigma_x \otimes\sigma_z\} \\
- \hat{p}_x \{\sigma_y \otimes\sigma_0\} +  \epsilon g/r\{\sigma_0 \otimes\sigma_x\}.\phantom{abc}
\end{eqnarray}
We can now define an antilinear operator $\hat{T} \equiv \{ \sigma_0 \otimes \sigma_y \}K$, where $K$ represents complex conjugation. It can easily be checked that $\hat{T}^2 = -1$. The definition of this operation, $\hat{T}$, is very similar to that of time-reversal. However, it is not the actual time reversal operator for this problem. Time reversal, in fact, squares to $+1$ in this case. 
We consider the transformation of the Hamiltonian under $\hat{T}$,
\begin{eqnarray}
\nonumber H_{\mathbf{K},\mathbf{K}'}  &\rightarrow& \hat{T}^{-1} H_{\mathbf{K},\mathbf{K}'} \hat{T}  = K \{ \sigma_0 \otimes \sigma_y \} H_{\mathbf{K},\mathbf{K}'}
 \{ \sigma_0 \otimes \sigma_y \}K\\
\nonumber  &=& g/r \{\sigma_0 \otimes \sigma_0 \}- \lambda \{\sigma_z \otimes \sigma_0\} + \hat{p}_y \{\sigma_x \otimes\sigma_z\} \\
&-& \hat{p}_x \{\sigma_y \otimes\sigma_0\} -  \epsilon g/r\{\sigma_0 \otimes\sigma_x\}.\phantom{abc}
\end{eqnarray}
We see that this is an approximate symmetry as the Hamiltonian is invariant except for the term proportional to $\epsilon g/r$ which is numerically very small. 
As we have an (approximate) antilinear symmetry of the Hamiltonian, Kramer's theorem applies and guarantees that the eigenstates will be degenerate. 

This approximate symmetry is an artefact of the Coulomb impurity potential which guarantees that the symmetry breaking term is very small. For instance, a $\delta$-function impurity potential does not give rise to degenerate eigenvalues. It can be easily seen that this argument does not apply for the non-trivial case wherein the mass term takes the form $- \lambda \{\sigma_z \otimes \sigma_z\} $ and is not invariant under the action of $\hat{T}$.

\section{Non-trivial case -  probability density in bound state wavefunction}
\renewcommand{\theequation}{B\arabic{equation}}
\renewcommand{\thefigure}{B\arabic{figure}}
\setcounter{equation}{0}
\setcounter{figure}{0}

As we are primarily interested in the non-trivial case, we calculate the bound state probability density for this case. The spinor wavefunction for a bound state is $\psi(\mathbf{r}) \sim \mathcal{A} \Psi_{\mathbf{K},j,\beta} e^{i\bf{K}.\bf{r}} + \mathcal{B}  \Phi_{\mathbf{K'},j',\beta'} e^{-i\bf{K}.\bf{r}}$, with the single-valley wavefunctions given in Eqs.~\ref{eq.psiphi} of the main text.
For the highest bound state, we choose $(j=1/2,\beta=0)$ and $(j'=1/2,\beta'=0)$; these quantum numbers correspond to the highest bound state in each valley. Furthermore, we choose $\mathcal{A}=\mathcal{B}$ to make the wavefunction valley-symmetric, as this choice gives good agreement with our numerical results. The obtained spinor wavefunction is
\begin{equation}
\psi(r,\theta) = \left(
\begin{array}{c}
a(r) e^{i \mathbf{K}\cdot \mathbf{r} } + c(r) e^{- i \mathbf{K}\cdot \mathbf{r}} \\
b(r) e^{i\theta+ i \mathbf{K}\cdot \mathbf{r}} + d(r) e^{-i\theta- i \mathbf{K}\cdot \mathbf{r}}  \end{array}
\right).
\label{eq.psij}
\end{equation}
We normalize this wavefunction by demanding $\int r dr d\theta \{ \psi^\dagger \psi \}= 1$. This fixes the total probability, adding contributions from both sublattices, to be unity. This fixes the overall normalization constant. At the single valley level, the norm can be computed analytically\cite{Dong2003}. However, in our case, we do not have a closed form for the norm as the inter-valley terms cannot be integrated analytically. As explained in the main text, when comparing with numerical results, we introduce an overall fitting parameter to account for normalization.

The probability density in the bound state is given by $\rho=\psi^{\dagger}\psi$, the sum of densities on both sublattices. Using the linear combination form of $\psi$ in Eq.~\ref{eq.psij} above, we obtain
\begin{eqnarray}
\nonumber \rho(r,\theta)&=& a^2 (r)+b^2 (r)+c^2 (r)+d^2 (r)\\
\nonumber &+& 2a(r)c(r)\cos{\{ 2 \mathbf{K}\cdot \mathbf{r} \}}\\
&+& 2b(r)d(r)\cos{\{ 2\mathbf{K}\cdot \mathbf{r}+2\theta\}}.
\label{eq.genprobdensity}
\end{eqnarray}
This is the probability density in the highest bound state -- given the parameters of the problem, $g$ and $t'$, we can substitute the explicit wavefunctions given in Eq.~\ref{eq.analyticalformofwavefunction} above. 
This leads to the density expression given in Eq.~\ref{eq.density} of the main text. 

The cosine terms arise from inter-valley scattering. Strictly speaking, as we are only interested in coarse-grained density, we may ignore these 
 terms which give oscillations with wavelength comparable to the lattice constant. However, we retain them as the numerics also shows comparable oscillations (albeit with oscillations occurring between sublattices).

We have obtained the density in the highest bound state by choosing  $(j=j'=1/2,\beta=\beta'=0,\mathcal{A}=\mathcal{B})$ as explained above. For lower bound states, it is not easy to find the $(j,j',\beta,\beta')$ quantum numbers associated with a particular state obtained from the numerics. 

\section{Current expression for bound state}
Proceeding in the same way as in the above discussion on probability density, we find the current using the bond current expression $
{\mathcal{J}}_{i,A}^{j,B}  = 
 it \{\psi^\ast_{A}(i)\psi_{B}(j)-\psi^\ast_{B}(j)\psi_{A}(i) \}$. 
 We reexpress the site positions using $\delta$, the bond vector, and $\mathbf{R}$, the distance of the bond center from the impurity, so that $\mathbf{r}_i = \mathbf{R} - \delta/2$ and $\mathbf{r}_j = \mathbf{R}+\delta/2$. We take $\mathbf{R}$ to be along the horizontal direction and $\delta$ to be vertical - see inset to Fig.~S1. We further assume $\vert\mathbf{R}\vert \gg \delta$, i.e., we focus on bonds that are far from the impurity.  
This allows us to approximate $a(\mathbf{R}\pm\mathbf{\delta}/2)\approx a(\mathbf{R})$ and $\theta_{\mathbf{R} \pm \delta/2} \approx \theta_\mathbf{R}$. Using the analytic spinor wavefunctions of Eq.~\ref{eq.psij} above, we obtain the expression for the bond current given in Eq.~\ref{eq.J_boundstate} of the main text.

\bibliographystyle{apsrev4-1} 
\bibliography{Haldane_coulomb}
\end{document}